# Identifying Quality Mersenne Twister Streams
# For Parallel Stochastic Simulations

Benjamin Antunes – Claude Mazel – David R.C. Hill

{benjamin.antunes, claude.mazel, david.hill}@uca.fr
Université Clermont Auvergne, CNRS, Clermont Auvergne INP,
Mines Saint-Etienne, LIMOS,
63000 Clermont-Ferrand, FRANCE

**Abstract**

The Mersenne Twister (MT) is a pseudo-random number generator (PRNG) widely used in High Performance Computing for parallel stochastic simulations. We aim to assess the quality of common parallelization techniques used to generate large streams of MT pseudo-random numbers. We compare three techniques: sequence splitting, random spacing and MT indexed sequence. The TestU01 Big Crush battery is used to evaluate the quality of 4096 streams for each technique on three different hardware configurations. Surprisingly, all techniques exhibited almost 30% of defects with no technique showing better quality than the others. While all 106 Big Crush tests showed failures, the failure rate was limited to a small number of tests (maximum of 6 tests failed per stream, resulting in over 94% success rate). Thanks to 33 CPU years, high-quality streams identified are given. They can be used for sensitive parallel simulations such as nuclear medicine and precise high-energy physics applications.

**1. Introduction**

Parallel stochastic simulation is widely used in many domains: High Energy Physics, transports, climate, finance, epidemiological modeling and so on. As for sequential applications, statistically sound streams of numbers are mandatory for the quality of the results (Hellekalek 1998; Maigne et al. 2004; Click et al. 2011). Deep knowledge about the parallelization of random numbers is not well spread and this sometimes leads to a bad usage of parallel pseudo-random number streams, impacting not only the quality of the results but also the reproducibility of the research made (Hill 2019). Pseudo-random number generators (PRNGs) are deterministic and repeatable from an initial status (or seed for old generators); they give the same random stream for the same initial state, this is essential for debugging. In High Performance Computing (HPC), we often parallelize stochastic simulations and most of them are embarrassingly parallel; they occupy up to 80% of the Worldwide LHC Computing Grid (WLCG) (Boyer 2022). For embarrassingly parallel Monte Carlo, we need independent random number streams for each core or threads which executes independent computing (without communications). Several good PRNGs can be used for parallel computing (Hill et al. 2013; L'Ecuyer et al. 2017). In this paper, we work on Mersenne Twister (MT) (Matsumoto and Nishimura 1998). It is a well-spread generator, present in many standard libraries of various languages used for HPC: C, C++, FORTRAN and many other languages. MT is also known for possible statistical flaws that makes it not suitable for cryptographic purposes. Some top PRNGs like MRG32k3a have better results to statistical tests and also propose a fine API (Application Programming Interface) for multiple parallel streams (L'Ecuyer 1999). However, MT is much faster than MRG32k3a, up to 19x in C on the latest x86 server processors when both codes are compiled with optimization flags (Hill 2022). If

MRG32k3a is slightly statistically better, in practice MT is often the most used PRNG for intensive stochastic computing where the time assigned to random drawings is important compared to the other part of the computing (Dao et al. 2014). In addition, the fact that MT has become a well-known "standard" increased our motivation to focus on this generator. To obtain independent parallel stochastic streams with this PRNG, several methods exist, they will be discussed later. We could not find in the literature information about the quality of streams depending on the parallelization technique used to "split" the sequence of an MT generator. In this paper, we test 4096 MT statuses produced with different parallelization techniques: Sequence Splitting, Random Spacing and MT Indexed Sequence (Hill et al. 2013). We applied the TestU01 Big Crush test battery to the parallel streams corresponding to all the produced statuses (L'Ecuyer and Simard 2007). A test result is considered a failure if the p-value is less than $10^{-10}$ or greater than $1-10^{-10}$; thus the confidence level is $1 - 2 \times 10^{-10}$. We ran this test (drawings of integer and real numbers) on 3 different hardware configurations, in order to assess the repeatability of the obtained results and to challenge the reproducibility of TestU01. The two aims of this paper are thus to check the quality of the random number streams depending on the parallelization technique, and to check the reproducibility of TestU01 for integer and real random numbers on different systems. We will first present the parallelization techniques used with good PRNGs with a short history of PRNG statistical tests. Next, we present our experiment, and how we managed to achieve 4096 parallel testing with 3 different random number parallelization techniques on 3 different hardware configurations, testing integer and real random numbers. Then, we add a short discussion about the terminology used around the notion of reproducibility as it recently changed in 2020 according to ACM. Finally, we show and discuss the results of this study about the quality of MT streams depending on the generator parallelization technique: does this quality depend on the parallelization method? All scripts, codes, data and high quality streams identified are open source, and available for researchers at https://gitlab.isima.fr/beantunes/testu01repro-and-mtstatuses.

## 2. Random Numbers and Their Parallelization for Simulations

### 2.1 A Short List of Good PRNGs

For stochastic simulations in HPC, using PRNGs remains the mainstream choice since repeatability is essential for debugging. True random numbers are still slow to produce and should be stored to retrace the simulation execution whether for debugging or for understanding discoveries. HPC simulations in High Energy Physics or nuclear medicine require thousands of experiments, some of them need up to $10^{12}$ random numbers for a single replicate and thousands of replicates may be required to improve statistics. For such intensive computing, even very fast storage of True random numbers cannot be considered. Quasi-random numbers is an option used for specific applications: integral computations, finance, … They have limitations when used in high dimensions if not improved (Sobol et al. 2011). Quantum computing, based on intrinsic randomness of quantum mechanics, will be interesting and available in a middle term to simulate quantum phenomena in a reasonable time (Feynmann 1982; Cluzel et al. 2019). In the meantime, PRNGs are the most efficient technique for the majority of applications and statistically sound generators are available for more than two decades. Streams of random numbers are produced deterministically, and the correlation proof is the source code of the random generator itself. However, when this code is a high quality model of randomness, the actual statistical tests can't find a correlation between the numbers drawn; this is why they are called commonly "random numbers". Maybe in some years, someone will create new statistical tests that will find that our current top PRNGs are doing bad. A full review of the best modern pseudo random number generators might require an entire paper, so we present here a selection of top PRNGs, some of them being widely used in HPC like the Mersenne Twister. In the paper of

Bhattacharjee and Das (2022), the most recent state of the art for PRNGs, authors describe some of the previous PRNGs, but lacks some of them. We do not want to champion a particular generator, because they all have their strengths and weaknesses. In addition, for critical applications, it is interesting to run the same simulation with various generators coming from a different 'family' (i.e., designed with different techniques) and then see if we observe a stochastic variability in the final results. Here is our short list for simulation and high performance computing:

Philox and Threefry were proposed by Salmon and his colleagues at the 2011 Supercomputing Conference (Salmon et al. 2011). They rely on cryptographic techniques like AES (Daemen and Rijmen 2001). They propose a sound and easy parameterization technique to solve the problem of distributing "independent" stochastic streams within parallel applications. These PRNGs do not appear in the latest state of the art survey about good PRNGs (Bhattacharjee and Das 2022). As observed by Mascagni and Hill, the implementation is sometimes not repeatable (Hill 2015). Because of its cryptographic basis, it is a slow generator, but statistically this kind of generators are very good.

MRG32k3a is a combined recursive pseudo random number generator selected by L'Ecuyer with brute force to meet the most stringent statistical tests (L'Ecuyer 1999). With MRG32k3a, it is also very easy to obtain a high number of parallel streams which is useful for parallel computing. However, when you need a huge amount of random numbers for intensive parallel computing, the fact is that this generator is 19 times slower than MT for its C/C++ version and this refrains to use it for intensive computing. MRG31k3p is also a sound generator from the same family.

WELL is a PRNG proposed by Panneton, L'Ecuyer and Matsumoto (Panneton et al. 2006). Like the famous Mersenne Twister, it is deriving from linear feedback shift registers (LFSR). This generator was presented to be an improvement of Mersenne Twister, which is true from a statistical point of view. However, Bhattacharjee and Das (2022) still ranks this PRNG worse than MT, because it is more seed dependent. Parallelization techniques were not proposed with this generator and in addition it is much less used (and spread) than MT.

PCG is the most recent PRNG of our short list. It was created in 2014 by O'Neill (2014) who states that it has better statistical properties than other generators. The Numpy documentation champions this generator as the best to use for general purposes. This generator is ranked between the 32 and 64 bits version of the Mersenne Twister in the latest survey of Bhattacharjee and Das (2022).

Mersenne Twister was proposed in 1998 by Matsumoto and Nishimura (1998), and a 2002 version updated the initialization scheme. This generator became quickly known because of its huge period of $2^{19337}$ - 1. It was the first of a family of generators, some designed for GPUs and Field Programmable Gate Array (FPGAs). The SFMT version uses the vector possibilities of modern processors (Single Instruction Multiple Data), and is faster than the original MT. It has better statistical properties and proposes an even larger period up to $2^{216091} - 1$, but it is much less known than the original MT (Saito and Matsumoto 2008). We have to remember that this family of generators are not suitable for cryptographic purposes.

Our list is not exhaustive and even if we do not want to champion a generator among the others for the previously exposed reason, it is interesting to mention the ranking achieved in the latest survey by Bhattacharjee and Das (2022). They rank the MT family at the 6 first places among 30 generators. SFMT being at the first two places in its 64 and 32 bits versions. The "original" MT holds the next places (4th and 5th). The 3rd place is hold by an elementary Cellular Automata with an autopletic rule (n°30) (Wolfram 1986), an interesting but very slow generator. Since we want to identify a set of high quality statuses which might help in sensitive applications requiring intensive stochastic computing (Maigne et al. 2004; Bernal et al. 2015; Boyer et al. 2022), we need specific statistical tests.

## 2.2 Statistical Tests for PRNGs

To measure the quality of a PRNG, we use statistical tests to separate good PRNGs from bad ones. We provide here an historical short list of tests ending by the best statistical suite we know:

Knuth tests: in his second volume of "The art of computer programming", Donald Knuth proposed a first set of statistical tests for PRNGs; they are still interesting to learn and cite (Knuth 1999).

Die hard: Marsaglia published in 1996 a small test suite composed of 15 tests. Die Hard source code is not accessible anymore; in the reference section we give a "wayback machine" website reference to reach his historical code (Marsaglia 1996).

Die Harder: Brown and his australian colleagues continued the Marsaglia pun by proposing a more recent update of tests as open-source software (Brown et al. 2013).

NIST – STS: The NIST Statistical Test Suite for random and pseudorandom number generators is the reference for cryptographic applications (Rukhin et al. 2010).

TestU01: L'Ecuyer and Simard proposed an open source library for empirical testing of random number generators. TestU01 is a very complete test suite with different levels of testing (Small Crush, Crush, Big Crush…) (L'Ecuyer and Simard 2007).

L'Ecuyer and Matsumoto also proposed some individual tests, and Bhattacharjee and Das (2022) are also using lattice tests and space–time diagram tests providing graphical outputs. In this paper, we have selected the TestU01 Big Crush test battery. We will use TestU01 tests for either sequences of uniform random numbers in [0,1] (as real numbers) or bit sequences (as integers). It is currently the hardest test battery to our knowledge, assessing streams with hundreds of billions of random numbers and thus, it meets our requirements for high performance stochastic computing.

## 2.3 Parallelization Methods for PRNGs

Parallel stochastic simulations are used in a lot of different science fields, where not all scientists have a deep knowledge about computer science. When properly designed, and when we remember that PRNGs are deterministic, it is possible to achieve repeatable parallel stochastic simulations. To gain confidence in the parallel execution, some research domains want to be able to compare the results obtained (at small scales) by a sequential execution and its parallel equivalent one. Once the results are satisfactory, we can run at larger scales. This approach has to be carefully designed as specified in the paper of Hill (2015). The sequential version must be prepared with a parallel version in mind with as many independent random streams as required for the corresponding parallel version. For instance, instead of using a single parallel stream with a loop for replicates, each replicate will use a different stream even in the sequential program. The purpose is to be able to trace each independent stochastic parts in the sequential and parallel execution. Archiving all the statuses used allow repeatable parallel executions. At small scales, the same statuses being used in the sequential version, we expect to obtain the same results that we have with the parallel execution. Different techniques are available to distribute independent random streams, they are presented by Hill et al. (2013). Mainly, there are two different approaches: the partitioning of a single generator stream and the generation of many parameterized generators supposed independent. With a parameterization approach, it is possible to create different smaller Mersenne Twisters (Matsumoto and Nishimura 2000). This approach has been tested by Reuillon (2008) on the original MT and by Passerat-Palmbach et al. (2011) for the GPU version (MTGP). It was worth testing as it showed limits that have been reproduced and quickly corrected by Makoto and his very responsive team always sensitive to the quality of a scientific work. Approaches which partitions a single generator are also used for parallel computing. In this paper, we did not retain the leap frog technique, nor the central server approach, the latter being not interesting for reproducible science with a different number of processing elements. The leap frog approach was not considered for MT, since the underlying jump ahead technique is much too slow for high performance computing (3 orders of magnitude slower than what we can achieve with MRG32k3a). We propose hereafter a list of the common techniques, that will be used in this paper:

Sequence splitting: This approach supposes that you can easily identify the maximum bound of random numbers required for all parallel processing elements. You can then decide an upper limit with a large margin in order to give sub streams to each processing elements. Splitting a large stream into equally spaced substreams supposes that you are able to archive the 'real status' of your PRNG to start each substream at the same position, this status is more complex than single seed used for old generators. When the size of the substreams is selected properly, they do not overlap. For example, if your simulation elements need at most $10^{10}$ random numbers, you must space each stream of at least $10^{10} + 1$, however a good habit is to be cautious by adding at least an order of magnitude for the spacing.

Random spacing: This approach generates random statuses suited for the generator. The risk might be that the PRNG is not well initialized. We know from previous work that the initialization can impact the quality of the random streams. Matsumoto et al. (2007) demonstrated that the majority of modern random number generators have correlations in their streams depending on the initialization state. Among the 58 generators in the GNU Scientific Library, 40 of them showed problems with careless initialization schemes. We also have to be careful about the problem of overlapping streams. In the paper of Hill et al. (2013), authors find that with a period like $2^{61}$, the probability of overlapping starts to be noticeable (0.43%), whereas it is close to 0 with huge periods like the one proposed by the MT family.

MT indexed sequence: When initializing with a simple seed in Mersenne Twister (with init_genrand()), the code is applying a randomization from the long type seed value, and fills the full mt[] tab with these values. At the end, it is like MT has a variant of random spacing already implemented and it generates a full 2.5 kB MT status from a simple seed.

Much more details can be found in the paper of Hill et al. (2013). Since the period of MT is incredibly huge, "independent" streams are easily obtained. We will use the 3 parallelization techniques presented above since they are suited for repeatable high performances simulations with MT. The remaining questions that we will handle is: do we obtain quality substreams and does the parallelization technique impacts this quality? Currently, we could not find research papers providing this information, nor a database of tested MT statuses. We want to answer the previous questions, in a reproducible way.

## 3. Material and Method

### 3.1 Experimental Design of the Study
In order to determine the quality of generated Mersenne Twister statuses depending on the parallelization technique, and second to challenge TestU01 reproducibility, we have selected a set of hardware configurations and a set of parallelization methods. As said previously, we know for a long time that the quality of pseudo random number streams is a determinant factor for research using parallel stochastic simulations (Lazaro et al. 2005; Click et al. 2011). Here is the material we have at our disposal for testing purpose: we have 3 different configurations A, B and C. In the A configuration, we have a cluster with 6 nodes running on Linux (Ubuntu) with 2 AMD EPYC 7452 with 32 physical cores and 512 GB RAM. Each node has thus 64 physical cores. For the B configuration, we used with direct access a machine with 2 AMD EPYC 7763 with 64 physical cores and 512 GB of RAM. This machine has 128 physical cores. Third, for the C configuration, we used an SMP SGI UV Brain machine with 16 Intel(R) Xeon(R) CPU E7-8890 v4 2.20GHz (Broadwell), leading to 384 physical cores (and 12 TB of RAM). The compiler used on the three machines is 'gcc'. The 'gcc' version is different on each machine we used, respectively 9.4.0, 11.3.0 and 4.8.4 for the UV Brain. This led to obtain small differences in the executable binary checked with the Linux "diff" command. We are studying the original Mersenne Twister (its 32 bits version) since TestU01 is designed in 32 bits. In addition to the fact that MT is being ranked as one of the best current generators, it is by far the most known and spread modern generator. Even if some flaws are known, it has now been used for a long period of time without claims of observed failures in real applications. In addition, conversely to what is being said in Wikipedia, the original MT is already very fast, particularly when compiled with optimization flags (SFMT is twice faster but less known and used). We have in high consideration the other modern generators that we mentioned, but we wanted to justify why we choose to study this PRNG in our paper. The fact that it is very well spread among the scientific community and standard libraries is maybe the main argument of our choice. Even if we have elements showing that MT is not 'seed' dependent, experience in nuclear medicine showed that we need to verify the quality of the initialization statuses. Indeed, the real status of MT is not a small seed but an array of 624 long numbers plus an index. We generated 4096 MT statuses with 3 different methods: Sequence splitting, random spacing and the method specific to MT which produces a full 2.5 KB status from an unsigned long integer (init_genrand() function). The size of $2^{12}$ streams was retained for the Sequence splitting approach; with the 4K statuses, it is adapted to large clusters or to small partitions of supercomputers. These statuses, and the associated code to produce them are available on our Gitlab. For sequence splitting, we started MT with the example initialization, and we sequentially saved statuses after $10^{12}$ drawings. This method first takes a long time since we have to generate *spaceBetweenStatus* x *numberOfStatuses* numbers. In our case we had to generate $10^{12}$ x 4096 numbers (~$4.10^{15}$ numbers). For random spacing, we are using the genrand_int32() function of MT to fill the elements of an MT status (the mt[] unsigned long array, and to give a proper mti integer index). We can generate as many statuses as we want, and this method is really fast to produce statuses. As discussed previously, the probability of over-lapping is close to 0 with the huge MT period ($2^{19337}$). For the MT index sequence, we are just using the init_genrand() which accepts an unsigned long and saves a status. Achieving this in a loop from 0 to 4095 is producing the last set of statuses. Matsumoto and his team carefully implemented this function to avoid bad initializations (Matsumoto et al. 2007). This method is also very fast compared to the first one. The only benefits of the sequence splitting approach is that the corresponding streams do not overlap and when the statuses are already computed we can reuse them directly. On the three different types of machines, we put the 4096 MT statuses corresponding to each method. A "diff –r" Linux command ensures us that all statuses are the same for each method on each machine. This check of MT status files eliminates reproducibility problems that could come from this part of the experiment. Before using the generated statuses in parallel, we have to test their quality. The TestU01 code comes from L'Ecuyer website, and we use version 1.2.3. The Big Crush test battery consists in 106 tests and each test is considered as a success if its *p*-value falls between $10^{-10}$ and $1 - 10^{-10}$. All the testing were executed in parallel with the machines at our disposal. The Big Crush test battery is relatively compute intensive. It needs around 4h time on modern processors for status testing. We have 4096 parallel testing with 3 different random number parallelization techniques on 3 different machines, with a testing of bit sequences for integer and for real random numbers. It corresponds to 73,728 runs of 4 hours. A total of 294,912 hours, more than 33 CPU years of computing was performed in two months of computing on our servers and clusters. The time needed to setup the experiment is variable depending on the technique used. The generation of MT statuses with random and MT index techniques is negligible (700 milliseconds for 4096 statuses, including the writing of status files), while the generation with sequence splitting techniques can take several days depending on the quantity of statuses; 1 hour for each status spaced by $10^{12}$ numbers on modern processors.

**3.2 A Recent Evolution of the "Reproducibility" Terminology**

While reproducibility can appear as a simple notion, the fact is that we don't have a consensus definition among researchers and research fields. In Philosophy of Science, reproducibility is considered as a cornerstone of Science. But what is really the meaning of reproducibility? Here was the main terms and definitions before 2020, according to the ACM:

Repeatability: Same team, same experimental setup; Reproducibility: Different team, different experimental setup; Replicability: Different team, same experimental setup (ACM badges).

In these definitions, as Drummond said (Drummond 2009), reproducibility requires changes, replicability avoids it. Reproducibility was the fact that a different team applying a different method or setup for the same scientific question, obtains the same scientific conclusions. This is strengthening the discovery. In the case of replicability, the goal was that a different team, using the first team article artifacts, can obtain the same results with a stated precision.

In literature, authors were sometimes using the word "reproducibility" to talk about "replicability", and sometimes it was the other way around. Advised by NISO (National Information Standards Organization), ACM changed their definitions after 2020, swapping terms between reproducibility and replicability for a better match with the practice of other research fields. Barba found which scientific fields used one definition or the other (Barba 2018). With this study, we realize that Computer Science was one of the rare fields to use the definition given at the beginning of this section. That was an argument to follow the NISO standardization advice; ACM then switched their reproducibility and replicability definitions. Here are the 2020, and current, definition for both terms:

*"Reproducibility (Different team, same experimental setup): The measurement can be obtained with stated precision by a different team using the same measurement procedure, the same measuring system, under the same operating conditions, in the same or a different location on multiple trials. For computational experiments, this means that an independent group can obtain the same result using the author's own artifacts.*

*Replicability (Different team, different experimental setup): The measurement can be obtained with stated precision by a different team, a different measuring system, in a different location on multiple trials. For computational experiments, this means that an independent group can obtain the same result using artifacts which they develop completely independently."* (ACM badges).

We can also observe that some authors, like us, active in the reproducibility research field in computer science, and that were using the previous definitions, like Stodden or Hinsen (Hinsen 2015; Rougier et al 2017), have also swapped their definitions of reproducibility and replicability; it now matches with the ACM 2020 update. Our approach is the same and in this paper, we will stick to the ACM definitions for "reproducibility" and "replicability". Increasing trust about a scientific conclusion with the fact that another team can obtain the same scientific conclusion is important for us.

The last term to discuss about is repeatability. It has led to less controversy than the two others. However, in computer science research papers, we sometimes find a confusion between repeatability and reproducibility. In our opinion, the definition of repeatability can still vary among scientific fields. In fact, in computer science, our machines are designed to be deterministic (except for quantum computing of course). We want to obtain run to run bitwise identical results on the same machine for the same program. This is absolutely essential for debugging and for the trust in the use of our deterministic computers. Repeatability is a real concern for researchers aware of program debugging, and this is particularly tough in the high performance computing world. Do we obtain the same weather forecast when we rerun the same experiment on the same supercomputer or cluster? This is an important question not so obvious since a decade. Ensuring reliable parallel debugging requires repeatability with bitwise identical results. And in this sense, we do not fully agree with the ACM definition which adds that results are identical with a "stated precision". This is because the ACM definition of repeatability comes from the International Vocabulary of metrology. In our opinion, this definition is perfectly correct for other sciences than computer science, where we really need "bitwise identical results" in order to debug properly.

## 4. Results and Discussion

When parallelizing a stochastic application, each stream given to a processing element has to be a good 'sequential random sequence'. The good news is that we have a perfect repeatability on the 3 configurations. On the other side, even if we were aware that MT has some known flaws, we were surprised by our results. In Table 1, we observe that almost 30% of the MT statuses generated with the different techniques give random streams that are failing to more than the 2 tests where they were expected to fail (tests for cryptographic applications, 80 and 81 LinearComp). The first column gives the names of the parallelization technique (test with integers or real numbers), the next 3 columns give the results obtained in the 3 configurations we have tested (A: cluster, B: massive multicore server, C: Huge SMP).

Table 1: Numbers of statuses that failed more than the two LinearComp tests on the 3 different configurations with 3 different parallelization techniques.

| Hardware configuration | A (cluster) | B (massive multicore server) | C (Huge SMP) |
|---|---|---|---|
| GCC version | 9.4.0 | 11.3.0 | 4.8.4 |
| Random spacing (int) | 1185 / 4096 28.93% | 1185 / 4096 28.93% | 1185 / 4096 28.93% |
| Random spacing (real) | 1185 / 4096 28.93% | 1185 / 4096 28.93% | 1185 / 4096 28.93% |
| Sequence Splitting (int) | 1156 / 4096 28.22% | 1156 / 4096 28.22% | 1156 / 4096 28.22% |
| Sequence Splitting (real) | 1156 / 4096 28.22% | 1156 / 4096 28.22% | 1156 / 4096 28.22% |
| MT Index Seq. (int) | 1139 / 4096 27.8% | 1139 / 4096 27.8% | 1139 / 4096 27.8% |
| MT Index Seq. (real) | 1128 / 4096 27.53% | 1128 / 4096 27.53% | 1128 / 4096 27.53% |

A remark for Table 1: when comparing the streams of integers and reals, it appears that we have the same statistic, but the streams do not fail to the same tests, this will be discussed later (even if we have the same count).

Secondly, we notice that the parallelization method used has no clear impact on the quality of the random streams. The small differences we can see are not statistically significant even if we can now give a little ranking in favor of the specific indexing function developed specifically for MT to produce a full status. Even if we can obtain 70% of very good parallel random streams, we have to dig a little more the question for the remaining 30% of failures. A first suggestion is that researchers working with MT and needing to use parallelization of stochastic simulations should avoid the Sequence splitting technique to generate statuses, except if statuses are already computed and available. Indeed, this approach needs a long sequential pre-computing with the unrolling of the MT sequence and the saving of statuses at a fixed distance. The test was achieved with non-overlapping sequences spaced with $10^{12}$ random numbers. Another interesting point shown by this table is a good news for numerical science: TestU01 is nicely repeatable and reproducible on different platforms in our experience. In fact, for each parallelization technique, we obtained numerical reproducibility on the three different hardware and software configurations (with different processors and even different executable files since each configuration came with a different gcc version). In Table 2, we can see that each parallel stream that failed to more than the 2 markers (Linear Comp testing), does not fail to more than 6 tests (out of 106). This is to say that even if we have a large number of defects, the quality of the streams remains good with more than 94% of success out of 106 tests. The main point is that the Sequence splitting method needs a huge preparation time without obtaining better results than the Random spacing or MT Index sequence.

Table 2: Numbers of failed tests and frequency of failures on the 3 different configurations.

| Name | Number of failed tests | Corresponding frequency |
|---|---|---|
| MT Index Sequence Integer | 3 | 964 |
| | 4 | 156 |
| | 5 | 18 |
| | 6 | 1 |
| MT Index Sequence Real | 3 | 951 |
| | 4 | 158 |
| | 5 | 19 |
| | 6 | |
| Sequence splitting Integer | 3 | 971 |
| | 4 | 164 |
| | 5 | 21 |
| | 6 | |
| Sequence splitting Real | 3 | 971 |
| | 4 | 161 |
| | 5 | 24 |
| | 6 | |
| Random spacing Integer | 3 | 990 |
| | 4 | 172 |
| | 5 | 21 |
| | 6 | 2 |
| Random spacing Real | 3 | 987 |
| | 4 | 175 |
| | 5 | 21 |
| | 6 | 2 |

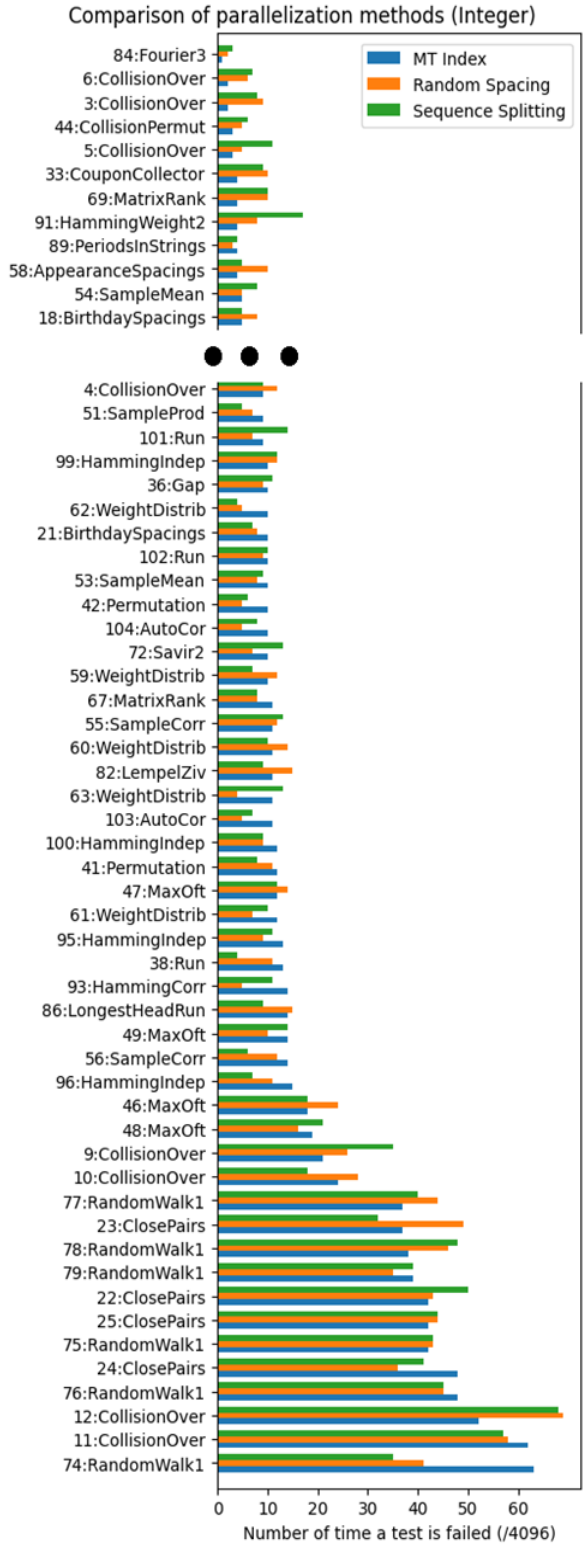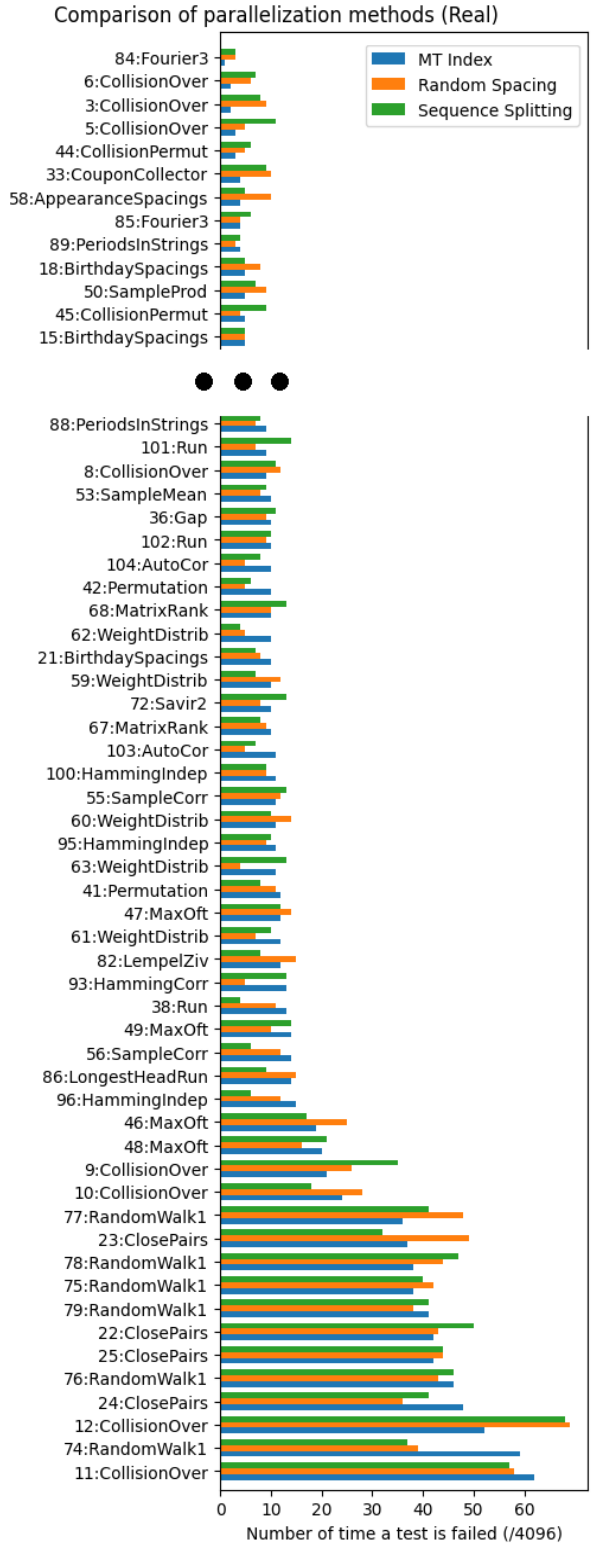

Figure 1: Frequency of test failure (from the TestU01 Big Crush battery) for integers and Figure 2: Frequency of test failure (from the TestU01 Big Crush battery) for real random numbers

We see in Table 2 that even if the integer and real random streams have the same number of failures, in the detail, they do not fail to the same tests. We cannot say that we have a better performance with one or the other, the quality of integer and real random number streams remain quite equivalent.

Figures 1 and 2 give results ordered by the frequency of failures for each test: all 106 tests are concerned! In these figures we truncate the list of tests to fit within a research paper page; full data are available with the artifacts. The three parallelization methods are compared for each test, and we have quite similar results. We can also notice that some tests detected a bigger number of defects.

We have made a selection of such tests and we list them in Table 3. This table only gives the tests with the biggest percentage of failure (around 1%) for the three parallelization techniques. When the test name is the same and the test number differs, it is in fact the same statistical test which is applied with different parameters as explained in the TestU01 user guide.

Table 3: Percentage of test failure to the TestU01
Big Crush battery with real numbers depending on the parallelization technique.

| Test numbers and names | MT Index Sequence failure % (on 4096) | Random Spacing failure % (on 4096) | Sequence Splitting failure % (on 4096) |
|---|---|---|---|
| 11:CollisionOver | 1.51% | 1.42% | 1.39% |
| 74:RandomWalk1 | 1.44% | 0.95% | 0.9% |
| 12:CollisionOver | 1.27% | 1.68% | 1.66% |
| 24:ClosePairs | 1.17% | 0.88% | 1.0% |
| 76:RandomWalk1 | 1.12% | 1.05% | 1.12% |
| 25:ClosePairs | 1.03% | 1.07% | 1.07% |
| 22:ClosePairs | 1.03% | 1.05% | 1.22% |
| 79:RandomWalk1 | 1.0% | 0.93% | 1.0% |
| 75:RandomWalk1 | 0.93% | 1.03% | 0.98% |
| 78:RandomWalk1 | 0.93% | 1.07% | 1.15% |
| 23:ClosePairs | 0.9% | 1.2% | 0.78% |
| 77:RandomWalk1 | 0.88% | 1.17% | 1.0% |

As we can see, results are pretty similar for each generation method. However, sequence splitting is the only one that takes much more time to generate statuses since it needs a pre-computing phase. Random spacing and Index sequence are much faster to generate these statuses. In fact, Sequence splitting needs to generate all numbers ($10^{12}$ x number of statuses), while the two others simply need to generate 624 values multiplied by the number of statuses. Between the two previous techniques, we can see a slightly better result for the method proposed by Matsumoto and Nishimura. From these facts, we would recommend using the index sequence technique to obtain 'independent' parallel sequences of MT numbers. However, the sequence splitting is the only method guaranteeing non-overlapping if we respect the size of the generated sequences.

The limitations of this study comes from the fact that we cannot test a huge amount of MT statuses due to the computing power needed to achieve all the testing. Another limitation is that a 64 bits version of TestU01 is not available.

## 5. Conclusion

In real applications where we need parallel random numbers, the quality of the parallelization technique can significantly impact the results (Maigne et al. 2004; Lazzaro et al. 2005; Hill et al. 2013).

In this paper, we studied the quality of Mersenne Twister streams with 3 parallelization techniques of the same generator: Random spacing, MT Indexed sequence and Sequence splitting. We achieved the tests on three different platforms for reproducibility purposes and we saw that we could not champion a technique among. Results show a very slight differences in favor of the MT specific technique. As Sequence Splitting is taking way more preparation time than the two others with the sequential computing of the initial statuses before using them in parallel, we suggest that its usage should be avoided, except when you already have at your disposal fine initial statuses (when the corresponding stream have been tested). The repeatability of our TestU01 experiments with different hardware and software showed that the produced results are robust. Our results show that 30% of the MT random number streams fail more than the two expected check tests (80 and 81 LinearComp). Since high quality streams are mandatory for reliable parallel stochastic simulations, the artifact of this paper provides: information and code to quickly generate quality statuses for MT and also all the statuses corresponding to high quality streams. In addition, to this set of "good statuses", we plan to continue assessing the reproducibility of TestU01, testing more features of the software library, on more different platforms.

## Acknowledgments

Mr. Antunes' thesis work is funded by the French MESRI (Ministry of Higher Education, Research and Innovation). Computations have been performed on the supercomputer facilities of the Mesocentre Clermont Auvergne.

## References

Barba, L. A. 2018. "Terminologies for Reproducible Research". *arXiv preprint arXiv:1802.03311*.

Bernal, M. A., M. C. Bordage, J. M. C. Brown et al. 2015. "Track Structure Modeling in Liquid Water: A Review of the Geant4-DNA Very Low Energy Extension of the Geant4 Monte Carlo Simulation Toolkit". *Physica Medica* 31(8):861-874.

Bhattacharjee, K., and S. Das. 2022. "A Search for Good Pseudo-Random Number Generators. Survey and Empirical Studies". *Computer Science Review* 45:100471

Boyer, A., C. Haen, F. Stagni, and D. Hill. 2022. "DIRAC Site Director: Improving Pilot-Job Provisioning on Grid Resources". *Future Generation Computer Systems* 133:23-38.

Brown, R. G., D. Eddelbuettel, and D. Bauer. 2013. "Dieharder: A Random Number Test Suite. Open Source Software Library, Under Development".
https://webhome.phy.duke.edu/~rgb/General/dieharder.php, accessed 05[th] August 2023.

Click, T. H., A. Lui, and G. A. Kaminski. 2011. "Quality of Random Number Generators Significantly Affects Results of Monte Carlo Simulations for Organic and Biological Systems". *Journal of Computational Chemistry* 32(3):513-524.

Cluzel, T., C. Mazel and D. Hill. 2019. "Quantum Computing: a Short Introduction". LIMOS UMR CNRS 6158, Research Report RR-2019-03778746, Université Clermont Auvergne, France, https://hal.uca.fr/hal-03778746

Daemen, J., and V. Rijmen. 2001. "Reijndael: The Advanced Encryption Standard". *Dr. Dobb's Journal: Software Tools for the Professional Programmer* 26(3):137-139.

Dao, V. T., H. Q. Nguyen, L. Maigne, V. Breton, and D. R. C. Hill. 2014. "Numerical Reproducibility, Portability and Performance of Modern Pseudo Random Number Generators: Preliminary Study for Parallel Stochastic Simulations Using Hybrid Xeon Phi Computing Processors". In *Proceedings of the European Simulation and modelling conference*, Porto, Portugal, 80-87.

Drummond, C. 2009. "Replicability Is Not Reproducibility: Nor Is It Good Science". In *Proceedings of the Evaluation Methods for Machine Learning Workshop at the 26th ICML*. Montreal, Canada: National Research Council of Canada.

Feynmann, R.. 1982. "Simulating Physics with Computers, Physics and Computation". *International Journal of Theoretical Physics* 21:467–488.

Haramoto, H., M. Matsumoto, T. Nishimura, F. Panneton and P. L'Ecuyer. 2008. "Efficient Jump Ahead For F_2-Linear Random Number Generators". *INFORMS Journal on Computing* 20:385–390.

Hellekalek, P. 1998. "Don't Trust Parallel Monte Carlo!" *ACM SIGSIM Simulation Digest* 28(1):82-89.

Hill, D. 2015. "Parallel Random Numbers, Simulation, And Reproducible Research". *Computing in Science & Engineering* 17(4):66-71.

Hill, D. 2019. "Repeatability Reproducibility, Computer Science and High Performance Computing: Stochastic Simulations Can Be Reproducible Too". In *Proceedings of the International Conference on High Performance Computing & Simulation (HPCS)*. IEEE. 322-323.

Hill, D. 2022. "Reproducibility of Simulations and High Performance Computing". In *Proceedings of the European Simulation and Modelling Conference*, ISEP Porto, ESM, Porto, Portugal, 5-9.

Hill, D., C. Mazel, J. Passerat-Palmbach et al. 2013. "Distribution of Random Streams for Simulation Practitioners". *Concurrency and Computation: Practice and Experience* 25(10):1427-1442.

Knuth, D. E. 1999. *The Art of Computer Programming. Numerical Algorithms,* Vol. 2. Addison Wesley/Pearson Education.

Hinsen, K. 2015. "Reproducibility, Replicability, and the Two Layers of Computational Science". https://khinsen.wordpress.com/2014/08/27/reproducibility-replicability-and-the-two-layers-of-computational-science/, accessed 05[th] August 2023.

L'Ecuyer, P. 1999. "Good Parameters and Implementations for Combined Multiple Recursive Random Number Generators". *Operations Research* 47(1):159-164.

L'Ecuyer, P. and R. Simard. 2007. "TestU01: AC Library for Empirical Testing of Random Number Generators". *ACM Transactions on Mathematical Software (TOMS)* 33(4):1-40.

L'Ecuyer, P., D. Munger, B. Oreshkin et al. 2017. "Random Numbers for Parallel Computers: Requirements and Methods, with Emphasis on Gpus". *Mathematics and Computers in Simulation* 135:3-17.

Lazaro, D., Z. El Bitar, V. Breton et al. 2005. "Fully 3D Monte Carlo Reconstruction in SPECT: A Feasibility Study". *Physics in Medicine & Biology* 50(16):3739.

Maigne, L., D. Hill, P. Calvat, et al. 2004. "Parallelization of Monte Carlo Simulations and Submission to a Grid Environment". *Parallel processing letters* 14(2):177-196.

Marsaglia, G. 1996. "DIEHARD: A Battery of Tests of Randomness". Source code of his CD on a way back machine website.
https://web.archive.org/web/20160125103112/http://stat.fsu.edu/pub/diehard/

Matsumoto, M., and T. Nishimura. 1998. "Mersenne twister: a 623-Dimensionally Equidistributed Uniform PRNG". *ACM Transactions on Modeling and Computer Simulation (TOMACS)* 8(1):3-30.

Matsumoto, M., and T. Nishimura. 2000. "Dynamic Creation of Pseudorandom Number Generators". *Monte Carlo and Quasi-Monte Carlo Methods Conference*. Springer. 56–69.

Matsumoto, M., I. Wada, A. Kuramoto et al. 2007. "Common Defects in Initialization of Pseudorandom Number Generators*". ACM Transactions on Modeling and Computer Simulation (TOMACS)* 17(4):15.

O'Neill, M. E. 2014. "PCG: A Family of Simple Fast Space-Efficient Statistically Good Algorithms for Random Number Generation". *ACM Transactions on Mathematical Software*.

Panneton, F., P. L'Ecuyer, and M. Matsumoto. 2006. "Improved Long-Period Generators Based on Linear Recurrences Modulo 2". *ACM Transactions on Mathematical Software (TOMS)* 32(1):1-16.

Passerat-Palmbach, J., C. Mazel, and D. Hill. 2011. "Pseudo-Random Number Generation on GP-GPU". In *Proceedings of the 25th ACM/IEEE/SCS Workshop on Principles of Advanced and Distributed Simulation (PADS 2011)*, 146-153.

Reuillon, R. 2008. "Testing 65536 Parallel Pseudo-Random Number Streams". *EGEE Grid User Forum*, Poster session, February 11-14th, Clermont-Ferrand.

Rougier, N. P., K. Hinsen, et al. 2017. "Sustainable Computational Science: the ReScience Initiative". *PeerJ Computer Science*, 3, 142.

Rukhin, A., J. Soto, J. Nechvatal et al. 2010. "A Statistical Test Suite for Random and Pseudorandom Number Generators for Cryptographic Applications", *Special Publication (NIST SP) - 800-22 Rev 1a*.

Saito, M., and M. Matsumoto. 2008. "SIMD-Oriented Fast Mersenne Twister: A 128-Bit Pseudorandom Number Generator". *Monte Carlo and Quasi-Monte Carlo Methods*. Springer Berlin Heidelberg, 607-622.

Salmon, J. K., M. A. Moraes, R. O. Dror, et al. 2011. "Parallel Random Numbers: as Easy as 1, 2, 3". In *Proceedings of 2011 International Conference for High Performance Computing, Networking, Storage and analysis,* 1-12.

Wolfram, S. 1986. "Random Sequence Generation by Cellular Automata", *Advances in Applied Mathematics* 7(2):123–169.

**Author Biographies:**

**Benjamin Antunes** is a Phd Student at Clermont Auvergne University (UCA). He holds a Master in Computer Science (head of the list). His thesis subject is about the reproducibility of numerical results in the context of high performance computing. He is espacially working on stochastic simultations. His email address is benjamin.antunes@uca.fr and his homepage is https://perso.isima.fr/~beantunes/.

**Claude Mazel** is Associate Professor since 1989 at Clermont Auvergne University (before 2021, Blaise Pascal university) and he joined the ISIMA Computer Science and Modelling Institute in 1993, where he managed the Computer-Aided Decision Processes, Information and Manufacturing Systems Department. His main scientific interests concern modelling, parallel stochastic simulations and statistical quality of their results. His email address is claude.mazel@uca.fr.

**David Hill** is doing his research at the French Centre for National Research (CNRS) in the LIMOS laboratory (UMR 6158). He earned his Ph.D. in 1993 and Research Director Habilitation in 2000 both from Blaise Pascal University and later became Vice President of this University (2008-2012). He is also past director of a French Regional Computing Center (CRRI) (2008-2010) and was appointed two times deputy director of the ISIMA Engineering Institute of Computer Science – part of Clermont Auvergne INP, #1 Technology Hub in Central France (2005-2007 ; 2018-2021). He is now Director of an international graduate track at Clermont Auvergne INP. Prof Hill has authored or co-authored more than 250 papers and has also published several scientific books. He recently supervised research at CERN in High Performance Computing. His email address is david.hill@uca.fr and his homepage is https://isima.fr/~hill/ .